\documentclass[12pt]{article}

\usepackage{latexsym}
\usepackage[colorlinks]{hyperref}
 \makeatletter

\def\be{\begin{equation}}
\def\ee{\end{equation}}
\def\ba{\begin{array}} \def\ea{\end{array}}
\def\bea{\begin{eqnarray}}
\def\eea{\end{eqnarray}}

\def\N{\mathbb N}

\usepackage{amsmath,amssymb}

\makeatletter
 \@addtoreset{equation}{section}

 \makeatother

\begin{document}
\thispagestyle{empty}

\vspace{20pt}

\baselineskip15pt

\begin{center}
{\Large\bf Nonlinear superconformal symmetry}

\vspace{10mm}

Mikhail Plyushchay $^{a,\,b}$

\vspace{5mm}
 ${}^a$ Departamento de F\'{\i}sica,
Universidad de Santiago de Chile,
Casilla 307, Santiago 2, Chile

\vspace{5mm}
${}^b$ Institute for High Energy Physics,
Protvino, Russia

\end{center}

\vspace{5mm}
\begin{abstract}
We discuss two different nonlinear generalizations of the
$osp(2|2)$ supersymmetry which arise in
superconformal mechanics and fermion-monopole
models.
\end{abstract}

\vskip 4cm
{\it Submitted to the Proceedings of the
International Workshop ``Supersymmetries and Quantum
Symmetries" (SQS'03, Dubna, Russia, 24-29 July 2003)}

\newpage

\section{Introduction}
Nonlinear supersymmetry \cite{AIS,P1,KP1,A1} is a
supersymmetric generalization
of a nonlinear symmetry characterized by a nonlinear algebra
of
integrals of motion. The best known examples of nonlinear
symmetry
are provided by the Kepler problem and by the planar
anisotropic
oscillator with a rational frequency ratio \cite{Walg}.
One of the simplest systems revealing nonlinear
supersymmetry
is a single-mode parabosonic oscillator system characterized
by the
superalgebra of the form
\be\label{ops}
[Q_+,Q_-]_{{}_+}=P_n(H),\quad
Q_\pm^2=0,\quad
[H,Q_\pm]=0
\ee
with a polynomial function $P_n(H)$ whose
order is
fixed by the order of a paraboson \cite{P1}.
In more general case of nonlinear supersymmetry
a polynomial may include dependence on other
even integrals of motion $I_k$,
\be\label{ngen}
[Q_+,Q_-]_{{}_+}=P_n(H, I_k), \quad
Q_\pm^2=0,\quad
[I_k,H]=[I_k,I_{k'}]=[I_k,Q_\pm]=[H,Q_\pm]=0,
\ee
playing the role
of the central charges of the nonlinear superalgebra
\cite{KP3}.

Nonlinear supersymmetry (\ref{ngen})
is a nonlinear generalization of the
linear $N=2$ supersymmetry (\ref{ops}) with $P_1(H)=H$.
The latter is a $(0+1)$-dimensional analog of the
super-Poincar\'e
symmetry. The natural question
that arises then is:
does there exist
a nonlinear generalization of superconformal symmetry
described in $(0+1)$-dimensional case
by the $osp(2|2)$ superalgebra?
The latter Lie superalgebra characterizes,
in particular,
the superconformal mechanics model
symmetry \cite{AP,FR,IKL}.

In this contribution,
based on the recent papers \cite{LP1,AnPl,LP2},
we shall show that the sought-for generalization
exists and, moreover, that it is a  hidden
symmetry of superconformal mechanics model when its
boson-fermion coupling constant takes integer
values. This can be compared
with the nonlinear symmetry of the planar anisotropic
oscillator appearing at the rational values of frequencies
ratio.
On the other hand, we shall see that
another, more simple nonlinear version of superconformal
symmetry characterizes the system of a charged fermion
in the field of the Dirac magnetic monopole.
The structure of the
nonlinear superconformal symmetry
of the fermion-monopole
system is close to the form
of nonlinear supersymmetry (\ref{ngen}).

\section{Nonlinear superconformal symmetry $osp(2|2)_n$}
Let us consider the
planar system of
a free spin-$1/2$ particle
described by the action
\begin{equation}\label{2d}
A=\int{\cal L}_0 dt,\quad
{\cal L}_0=\frac{1}{2}\dot{x}_i^2-
\frac{i}{2}\dot\xi_i\xi_i.
\end{equation}
It is characterized by the nontrivial Poisson-Dirac brackets
$\{x_i,p_j\}=\delta_{ij}$, $\{\xi_i,\xi_j\}=-i\delta_{ij}$,
$i,j=1,2$,
and by the integrals of motion $p_i$, $X_i=x_i-tp_i$
and $\xi_i$. Among all the functions of $p_i$, $X_i$ and
$\xi_i$,
there is a set of quadratic integrals
\begin{equation}\label{2dev}
H=\frac{1}{2}p_i^2,\quad
K=\frac{1}{2}X_i^2,\quad
D=\frac{1}{2}X_ip_i,\quad
\Sigma=-\frac{i}{2}\epsilon_{jk}\xi_j\xi_k,
\end{equation}
\be\label{tot}
J=L+\Sigma,
\ee
\begin{equation}\label{2dod}
Q_1=p_i\xi_i,\quad
Q_2=\epsilon_{ij}p_i\xi_j,\quad
S_1=X_i\xi_i,\quad
S_2=\epsilon_{ij}X_i\xi_j,
\end{equation}
where $\Sigma$ is the particle's spin, and
$
L=\epsilon_{ij}X_ip_j
$
is its angular momentum.
The even,
(\ref{2dev}), (\ref{tot}), and the odd,
(\ref{2dod}), integrals
form the superalgebra
(only the nontrivial Poisson bracket relations
are displayed)
\begin{equation}
\{D,H\}=H,\quad
\{D,K\}=-K,\quad
\{K,H\}=2D,
\label{dhk}
\end{equation}
\begin{equation}
\{D,Q_a\}=\frac{1}{2}Q_a,\quad
\{D,S_a\}=-\frac{1}{2}S_a,\quad
\{H,S_a\}=-Q_a,\quad
\{K,Q_a\}=S_a,
\label{hkqs}
\end{equation}
\begin{equation}
\{\Sigma,Q_a\}=\epsilon_{ab}Q_b,\quad
\{\Sigma,S_a\}=\epsilon_{ab}S_b,
\label{lsqs}
\end{equation}
\begin{equation}
\{Q_a,Q_b\}=-i\delta_{ab}2H,\quad
\{S_a,S_b\}=-i\delta_{ab}2K,\label{qsdj}
\ee
\be\label{qsj}
\{Q_a,S_b\}=-i\delta_{ab}2D -i\epsilon_{ab}(J+\Sigma).
\ee
The total angular momentum $J$
commutes with all other quadratic integrals
$H$, $K$, $D$, $\Sigma$, $Q_a$, $S_a$, $a=1,2$,
and the superalgebra
(\ref{dhk})--(\ref{qsdj}) is identified as the
$osp(2|2)\oplus u(1)$
with the $u(1)$ corresponding to the centre
$J$. The Hamiltonian reduction of the system to the surface
of the fixed value of $J$ given by the constraint
\begin{equation}\label{cons}
J-\alpha=0
\end{equation}
does not change the form of the superalgebra $osp(2|2)$,
replacing $J$ for a constant $\alpha$ in Eq. (\ref{qsj}).
Under such a reduction, the superconformal algebra
generators
take the form of the integrals of motion of the
superconformal mechanics
model \cite{AP,FR},
\begin{equation}\label{hscm}
H=\frac{1}{2}\left(p^2+
{\alpha(\alpha+2i\psi_1\psi_2)}q^{-2}\right),
\end{equation}
\begin{equation}
D=
\frac{1}{2}qp-tH,\qquad
K=
\frac{1}{2}q^2-2tD-t^2H,\qquad
\Sigma=-i\psi_1\psi_2,
\label{tdk}
\end{equation}
\begin{equation}
Q_a=p\psi_a+\frac{\alpha}{q}\epsilon_{ab}\psi_b,\qquad
S_a=q\psi_a-tQ_a,
\label{qs}
\end{equation}
where $q=\sqrt{x_i^2}$,
$p=n_ip_i$, $\psi_1=n_i\xi_i$,
$\psi_2=\epsilon_{ij}n_i\xi_j$,
$n_i= q^{-1} x_i$,
$\{q,p\}=1,$ $\{\psi_a,\psi_b\}=-i\delta_{a,b}$.

Now, let us generalize the constraint (\ref{cons})
for the constraint
\be\label{ncons}
{\cal J}_n-\alpha=0,\qquad {\cal J}_n\equiv L+n\Sigma,
\ee
where
$n=\N$.
The Lagrangian
\begin{equation}
{\cal L}_n={\cal L}_0-\frac{1}{2x_i^2}(\epsilon_{jk}x_j
\dot x_k+n\Sigma-\alpha)^2
\label{ln}
\end{equation}
with ${\cal L}_0$ given by Eq.
(\ref{2d})
generates constraint (\ref{ncons})
as the unique (primary) constraint
for the system with the canonical Hamiltonian
$H=\frac{1}{2}p_i^2$.
The quantities, gauge invariant
with respect to the action
of the constraint ${\cal J}_n$,
are identified as observables
of the system (\ref{ln}).
Defining the complex
variables
\be\label{xi+-}
X_\pm=\frac{1}{\sqrt{2}}
(X_1\pm iX_2),\quad
P_\pm=\frac{1}{\sqrt{2}}(p_1\pm ip_2),\quad
\xi_\pm=\frac{1}{\sqrt{2}}(\xi_1\pm i\xi_2)
\ee
with nontrivial Poisson bracket relations
$
\{X_+,P_-\}=\{X_-,P_+\}=1,
$
$\{\xi_+,\xi_-\}=-i$,
one finds the quadratic observables being the
integrals of motion of the system (\ref{ln}). These are
the ${\cal J}_n$ given by Eq. (\ref{ncons})
with $L=i(X_+P_--X_+P_-)$, and
\bea\label{hdk+}
&H=P_+P_-,\quad
K=X_+X_-,\quad
D=\frac{1}{2}(X_+P_-+P_+X_-),\quad
\Sigma=\xi_+\xi_-,&\\
\label{sn}
&S^+_{n,l}=2^{n/2}(i)^{n-l}
(P_-)^{n-l}(X_-)^l\xi_+,\quad
S^-_{n,l}=2^{n/2}(-i)^{n-l}(P_+)^{n-l}
(X_+)^l\xi_-&
\eea
with
$l=0,\ldots, n$.
At n=1 the odd observables
(\ref{sn})
are the linear combinations of
the odd integrals (\ref{2dod}).

On the surface of the constraint
(\ref{ncons}),
the relation
\begin{equation}
{\cal C}\equiv 4(KH-D^2)+2n\Sigma=\alpha^2
\label{cas}
\end{equation}
is valid, and  the quantity ${\cal C}$
commutes with all the set of the integrals
(\ref{hdk+}), (\ref{sn}).
The even integrals (\ref{hdk+}) form, as before, the Lie
algebra
$so(1,2)\oplus u(1)$. Then,
treating Eq. (\ref{ncons})
as the constraint that fixes the orbital angular momentum
$L$, and
taking into account the relation (\ref{cas}),
we find that on the surface (\ref{ncons})
the integrals (\ref{hdk+})
and (\ref{sn})
form the nonlinear superalgebra given
in addition to Eq. (\ref{dhk}) by the
following nontrivial Poisson bracket relations:
\begin{equation}\label{dsigs}
\left\{D,S^\pm_{n,l}\right\}=
\left(\frac{n}{2}-l\right)S^\pm_{n,l},\qquad
\{\Sigma,S^\pm_{n,l}\}=\mp iS^\pm_{n,l},
\end{equation}
\begin{equation}\label{hks}
\left\{H,S^\pm_{n,l}\right\}=
\pm ilS^\pm_{n,l-1},\qquad
\left\{K,S^\pm_{n,l}\right\}=
\pm i(n-l)S^\pm_{n,l+1},
\end{equation}
\begin{eqnarray}
\left\{ S_{n,m}^{+},S_{n,l}^{-}\right\}=
-i(2H)^{n-m}(2K)^{l}(\alpha -2iD)^{m-l}-i\Sigma
(2H)^{n-m-1}(2K)^{l-1}\times\nonumber\\
\left( \alpha
-2iD\right) ^{m-l}\left(
n\left( m-l\right) \left( \alpha -2iD\right) +4\alpha
l\left( n-m\right)
\right), \qquad m\geq l.
\label{slong}
\end{eqnarray}
The brackets between the odd integrals for the case
$m<l$ can be obtained from (\ref{slong})
by a complex conjugation.
The relations (\ref{dhk}), (\ref{dsigs})-(\ref{slong})
give a nonlinear generalization of the
superconformal algebra $osp(2|2)$
with the Casimir element (\ref{cas}).
In this nonlinear superconformal algebra,
denoted in ref. \cite{AnPl} as $osp(2|2)_n$,
the sets of odd generators
$S_{n,l}^+$ and $S_{n,l}$, $l=0,\ldots,n$,
form the two spin-$\frac n2$ representations
of the bosonic Lie subalgebra $so(1,2)\oplus u(1)$.

The quantum analogs
of the $osp(2|2)_n$ generators
(\ref{hdk+}), (\ref{sn})
are given by
set of the operators \cite{LP1}
\begin{equation}\label{hnq}
\hat H=\frac{1}{2}\left(
-\frac{\partial ^2}{\partial q^2}+\left(
a_n+b_n\sigma_3\right)\frac{1}{q^2}\right),
\end{equation}
\begin{equation}\label{dnq}
\hat D=-\frac{i}{2}\left(q\frac{\partial}{\partial q}
+\frac{1}{2}
\right)-\hat Ht,
\qquad
\hat K=\frac{1}{2 }q^2-2\hat Dt-
\hat Ht^2,\qquad
\hat \Sigma=\frac{1}{2}\sigma_3,
\end{equation}
\be\label{snq}
\hat S{}^+_{n,l}=\left(
q+it{\cal D}_{\alpha-n+1}
\right)
\left(
q+it{\cal D}_{\alpha-n+2}
\right)\ldots
\left(
q+it{\cal D}_{\alpha-n+l}
\right)
{\cal D}_{\alpha -n+l+1}\ldots
{\cal D}_\alpha
\sigma_+,\quad
\hat S{}^-_{n,l}=\left(
\hat S{}^+_{n,l}\right)^\dagger,
\ee
where $\sigma_+=\frac{1}{2}(\sigma_1+
i\sigma_2)$ and
\begin{equation}\label{anq}
a_n=\alpha_n^2+\frac{1}{4}(n^2-1),\quad
b_n=- n\alpha_n,\quad
\alpha_n=\alpha-\frac{1}{2}(n-1),\quad
{\cal D}_\gamma=\frac{\partial}{\partial
q}+\frac{\gamma}{q}.
\end{equation}
The second terms in $a_n$ and $\alpha_n$
in eq. (\ref{anq})
(proportional to $(n^2-1)$ and $(n-1)$)
include the quantum factors $\hbar^2$ and $\hbar$,
respectively,
while the term $\frac{\gamma}{q}$
in ${\cal D}_\gamma$
includes the factor $\hbar(=1)$.
These quantum corrections in the quantum
analogs of the corresponding classical quantities
can be obtained by the application of the reduction
procedure `first quantize and then reduce' to the system
(\ref{ln}) \cite{AnPl}.
This procedure fixes also the form of
the quantum analogs of the classical relations
(\ref{dhk}), (\ref{dsigs}), (\ref{hks}),
(\ref{slong}):
\begin{equation}
[\hat H,\hat K]=-2i\hat D,\quad
[\hat D,\hat H]=i\hat H,\quad
[\hat D,\hat K]=-i\hat K,
\label{qso}
\end{equation}
\begin{equation}
[\hat \Sigma,\hat S{}^\pm_{n,l}]=\pm \hat S{}^\pm_{n,l},
\quad
[\hat D,\hat S{}^\pm_{n,l}]=i\left(\frac{n}{2}-l\right)
\hat S{}^\pm_{n,l},
\label{sds}
\end{equation}
\begin{equation}
[\hat H,\hat S{}^\pm_{n,l}]=\mp l\hat S{}^\pm_{n,l-1},\quad
[\hat K,\hat S{}^\pm_{n,l}]=\mp (n-l)
\hat S{}^\pm_{n,l+1},
\label{hks1}
\end{equation}
\begin{eqnarray}
[\hat S{}^+_{n,m},\hat S{}^-_{n,l}]_{{}_+}=&&
\sum_{s=0}^{min(l,n-m)}2^ss!C^s_{n-m}C^s_{l}
\times (
(2\hat K)^{l-s}(2\hat H)^{n-m-s}
{\cal P}_{m-l+s}(-2i\hat D +c_s)
\Pi_+ +
\nonumber\\&&
(-1)^{m-l}(2\hat H)^{n-m-s}(2\hat K)^{l-s}
{\cal P}_{m-l+s}(
2i\hat D+d_s)
\Pi_-),
\label{ssquant}
\end{eqnarray}
where
$\Pi_\pm=\frac{1}{2}\pm \Sigma$,
$min(a,b)=a$ (or, $b$) when $a\leq b$
(or, $b\leq a$), $C^s_l=\frac{l!}{s!(l-s)!}$,
${\cal P}_k(z)$ is a polynomial of order
$k$,
\[
{\cal P}_0(z)=1,\qquad
{\cal P}_k(z)=z(z+2)\ldots (z+2(k-1)),\quad k>0,
\]
and
\[
c_s=\alpha
+\frac{3}{2}+n-2(m+s),\qquad
d_s=-\alpha +\frac{1}{2}+2(l-s).
\]
In (\ref{ssquant}) it is supposed $m\geq l$,
while the case corresponding to $m< l$ is obtained from
it by the Hermitian conjugation.
The quantum analog of the Casimir element
(\ref{cas})
of the
superconformal
algebra $osp(2|2)_n$  takes the value
\begin{equation}
\hat {\cal C}\equiv 2(\hat H\hat K+\hat K\hat H)-
4\hat D{}^2+2n\alpha_n \hat \Sigma  =
\alpha_n^2+\frac{1}{4}n^2-1.
\label{qcas}
\end{equation}

The obtained  nonlinear superconformal symmetry
$osp(2|2)_n$ is generated by
the four bosonic integrals
(\ref{hnq}), (\ref{dnq}),
which form the $so(1,2)\oplus u(1)$
Lie subalgebra,
and  by the $2(n+1)$ fermionic integrals
(\ref{snq})
constituting the two
spin-$\frac{n}{2}$
$so(1,2)$-representations
and anticommuting for the order
$n$ polynomials of the even generators.
In other words,
the found nonlinear generalization $osp(2|2)_n$
of the superconformal symmetry $osp(2|2)$
involves the
extension of the total number of the odd generators
from $4$ to $2(n+1)$.

\section{Hidden $osp(2|2)_n$ of
superconformal mechanics model}
Proceeding from the explicit form of the quantum
Hamiltonian (\ref{hnq}), one can find that
the equality
\be\label{hhnn}
\hat{H}_n^\alpha=\hat{H}_{n'}^{\alpha'}
\ee
takes place
for the
given values of $n$ and $n'\neq n$ when
the model parameter takes one of the four
corresponding
sets of
values
\bea
&\alpha=\alpha '=\frac{1}{2}(n+n '-1),\qquad
\alpha=-(\alpha '+1)=\frac{1}{2}(n- n' -1),&\label{aann1}\\
&\alpha -n=-(\alpha '+1)=\frac{1}{2}(n+n '+1),\qquad
\alpha-n=\alpha '=\frac{1}{2}(n- n' +1).&
\label{aann2}
\eea
The first two cases (\ref{aann1})
result in the following forms of the
Hamiltonian
\begin{equation}
\hat{H}=\frac{1}{2}\left(-\frac{d^2}{dq^2}+\frac{(n\pm
n')^2-1}
{4q^2}\mp \frac{nn'}{q^2}\Pi_+\right),
\label{hami1}
\end{equation}
where $\Pi_+=\frac{1}{2}(1+\sigma_3)$, and
the upper and lower signs correspond to the
first and second cases from (\ref{aann1}).
The Hamiltonian for the cases (\ref{aann2})
can be obtained from Eq. (\ref{hami1})
via the formal change
$n'\rightarrow n'\mp 2(n+1)$.
In particular case of $n'=1$, $n=2k$ and
$\alpha=\alpha'=k$, $k\in N$, corresponding to the first
relation
from (\ref{aann1}), the Hamiltonian takes the form
\begin{equation}
\hat{H}=\frac{1}{2}\left(-\frac{d^2}{dq^2}+{k}
(k- \sigma_3)q^{-2}\right),\quad
k\in \N.
\label{hamik}
\end{equation}
It is a direct quantum analog of the classical
superconformal mechanics Hamiltonian
(\ref{hscm}). Therefore, one can conclude that when the
boson-fermion
coupling constant takes integer values, $\alpha=k$,
in addition to the usual superconformal symmetry
of the order $n'=1$, the quantum system
(\ref{hamik}) possesses also the nonlinear
superconformal symmetry of the order $n=2k$,
the odd generators of which produce,
via the anticommutators with the
fermionic $n'=1$ superconformal symmetry generators,
the additional nontrivial
bosonic integrals of motion having a
nature of the half-integer degrees
of the odd order polynomials
of the $so(1,2)\times u(1)$ generators, see ref. \cite{LP1}
for the details.

\section{Fermion-monopole nonlinear superconformal symmetry}
Let us consider now the system of a charged fermion in the
field
of the Dirac monopole
described by the Hamiltonian
\begin{equation}
H=\frac{1}{2}P_i^2-eB_i{\cal S}_i
\label{hfm}
\end{equation}
with
$P_i=p_i-eA_i$,
$B_i=\epsilon_{ijk}\partial_jA_k=gx_i/|x|^3$,
$|x|=\sqrt{x_ix_i}$,
${\cal S}_j=-\frac{i}{2}\epsilon_{jkl}\xi_k\xi_l$,
and by the fundamental Poisson brackets
$\{x_i,p_j\}=\delta_{ij}$,
$\{\xi_j,\xi_k\}=-i\delta_{jk}$.
The Hamiltonian (\ref{hfm}) and the quantities
\begin{equation}
D=\frac{1}{2}X_iP_i+etB_i{\cal S}_i=\frac{1}{2}x_iP_i-tH,
\label{dfm}
\end{equation}
\begin{equation}
K=\frac{1}{2}X_i^2-et^2B_i{\cal S}_i=
\frac{1}{2}x_i^2-2tD-t^2H,
\label{kfm}
\end{equation}
where $X_i=x_i-tP_i$,
together with the full
angular momentum
$J_i$, given by the relations
\begin{equation}
J_i=L_i-\nu n_i+{\cal S}_i,\qquad
L_i=\epsilon_{ijk}x_jP_k,\qquad
n_i=\frac{x_i}{|x|},\qquad
\nu=eg,
\label{jil}
\end{equation}
constitute the set of the integrals
of motion generating the $so(1,2)\oplus so(3)$
symmetry \cite{Jackiw,mpmono}.
Since the quantization of the spin degrees of freedom
gives rise to the same two-dimensional
space associated with the Pauli matrices as in the
superconformal mechanics model,
it is rather natural to expect that the full symmetry
of the fermion-monopole system has to  have a nature of the
superconformal symmetry.
We shall show below that this is indeed so by exploiting the
analogy
with the two-dimensional description of the superconformal
mechanics model based on the Lagrangian (\ref{ln})
$n=1$.
For the purpose, we note that
the integrals of the `extended' superconformal model
(\ref{ln}) ($n=1$)
may be represented in a 3D form
if to introduce into the system the classical
odd Grassmann variable $\xi_3$
having the only nontrivial bracket
$\{\xi_3,\xi_3\}=-i$.
This will be the odd integral of motion $\Gamma$, whose
quantum analog
will coincide up to a numerical factor
with the even quantum operator
$\hat\Sigma=\frac{1}{2}\sigma_3$.
Being interpreted as odd operator, it will anticommute
with all the odd $osp(2|2)$ generators and will commute
with all its even generators, and so,
may be treated as a grading operator for
$osp(2|2)$.
Introducing the notations
$x_i=(x_1,x_2,0)$,
$p_i=(p_1,p_2,0)$,
$\xi_i=(\xi_1,\xi_2,\xi_3)$ and
$N_i=(0,0,1)$,
one can understand
the expressions for the
$so(1,2)$ generators from (\ref{2dev}) as
given by $3D$ scalar products,
whereas the integral $\Sigma$
together with the odd integrals (\ref{2dod}) and
$\Gamma=\xi_3$
can be represented in the 3D form as follows:
\begin{equation}
Q_1=p_i\xi_i,\quad
Q_2=\epsilon_{ijk}N_ip_j\xi_k,\quad
S_1=X_i\xi_i,\quad
S_2=\epsilon_{ijk}N_iX_j\xi_k,\quad
\Sigma=-\frac{i}{2}\epsilon_{ijk}N_i\xi_j\xi_k,\quad
\Gamma=N_i\xi_i.
\label{qnpsi}
\end{equation}
Then, defining
for $L_i^2\neq 0$
the vector
$$
N_i=
\left(
1-\frac{2\nu}{L_j^2}{\cal S}_kn_k\right)Y_i,\qquad
Y_i=L_i+
\frac{2}{3}{\cal S}_i,
$$
one can find that the quantities of the form (\ref{qnpsi})
with $p_i$ changed  for $P_i$ are the integrals
of motion of the fermion-monopole system if
the point $x_i=0$ is excluded from its
configuration space \cite{DHV,LP2}.
Using the classical relations $\xi_i{\cal S}_j=\frac{1}{3}
\delta_{ij}(\xi_k{\cal S}_k)$ and
${\cal S}_i{\cal S}_j=0$,
the integrals of motion additional to the
$so(1,2)\oplus su(2)$ generators
(\ref{hfm}), (\ref{dfm}),
(\ref{kfm}), (\ref{jil})
can be represented in the form
\be
Q_1=P_i\xi_i,\quad
Q_2={\cal P}_i\xi_i,\quad
S_1=X_i\xi_i,\quad
S_2={\cal X}_i\xi_i,\quad
\Sigma=L_i{\cal S}_i,\quad
\Gamma=L_i\xi_i+\frac{2}{3}{\cal S}_i\xi_i,
\label{qpqx}
\end{equation}
where
\begin{equation}
{\cal P}_i= \epsilon_{ijk}L_jP_k+\frac{2}{3}
\nu|x|^{-1}{\cal S}_i,
\qquad
{\cal X}_i=\epsilon_{ijk}
L_jX_k-\frac{2}{3}t\nu
|x|^{-1}{\cal S}_i.
\label{calpx}
\end{equation}
The odd integral $\Gamma$ satisfies the relation
\be\label{gamma}
\{\Gamma,\Gamma\}=-i{\cal J},\qquad {\cal J}=J_i^2-\nu^2,
\ee
and has zero Poisson brackets with all other
even and odd integrals of motion.
It is the classical analog of the Yano supercharge
found in \cite{MacHol}, see also refs.
\cite{Spec,PlFM,Horv}.

The even, $H$, $K$, $D$, $\Sigma$, $J_i$,
and odd, $Q_a$, $S_a$, $a=1,2$,
integrals of motion generate the superalgebra
similar to (\ref{dhk})--(\ref{qsj})
with the $J$ in Eq. (\ref{qsj}) changed for ${\cal J}$,
and
with the following relations
to be different from the corresponding $osp(2|2)$
relations:
\bea\label{qsj1}
&\{Q_2,Q_2\}=-2i{\cal J}H,\quad
\{S_2,S_2\}=-2i{\cal J}K,\quad
\{Q_2,S_2\}=-2i{\cal J}D,&\\
\label{qsj2}
&\{\Sigma,Q_2\}=-{\cal J}Q_1,\quad
\{\Sigma,S_2\}=-{\cal J}S_1.&
\eea
Therefore, one concludes that classically
the fermion-monopole
system possesses a symmetry which is a nonlinear
generalization of the superconformal symmetry
$osp(2|2)$ plus decoupled rotational symmetry $su(2)$
and the supersymmetry generated by the odd supercharge
$\Gamma$ being effectively `the square root' from
the central charge ${\cal J}$.
The central charge ${\cal J}$ appears additively
and multiplicatively in the generalized
$osp(2|2)$ relations, and in this respect
the nonlinear superconformal symmetry
of the fermion-monopole system has a structure
similar to the nonlinear superalgebraic structure
(\ref{ngen}).

The specific feature of the quantum analog of
the described nonlinear superconformal symmetry of the
fermion-monopole
system is encoded in the relations
\be\label{specq}
\hat{\Sigma}=\sqrt{\frac{\hbar}{2}}\hat\Gamma=
\frac{\hbar}{2}(\hat{L}_i\sigma_i+\hbar),\qquad
\hat{\Sigma}{}^2=\frac{\hbar}{4}\hat{\cal J},\qquad
\hat{\cal J}=\hat{J}_i^2-\nu^2+\frac{\hbar^2}{4},
\ee
where
$\nu$ is quantized,
$|\nu|=\hbar^2 k/2$, $k\in \N$,
and we have restored the quantum
constant.
These relations lead to the two consequences.
First, quantum mechanically
the odd integral $\hat{\Gamma}$,
being the grading
operator of nonlinear generalization of
$osp(2|2)\oplus su(2)$,
is not anymore independent from the even generator
$\hat{\Sigma}$ being different from it only in the
constant quantum factor. Second,
in representation where the
squared full angular momentum operator
is diagonal,
$\hat{J}_i^2=j(j+1)\hbar^2$,
$j+\frac 12=|\nu|+m$,
$m=0,1,2,...$,
we have $\hat{\cal J}=(|\nu|+m)^2-\nu^2$
\cite{mpmono,DHV}.
Then, following  ref. \cite{DHV}, one can show
that in the sector  $m=0$, corresponding
classically to the phase space surface
given by the equations $L_i=0$, ${\cal S}_jn_j=0$,
the symmetry of the system
is reduced to the conformal symmetry $so(1,2)$.

\section{Discussion and outlook}
To conclude, let us discuss some open
problems which deserve further attention.

For the integer values of the boson-fermion coupling
constant,
the quantum superconformal mechanics model possesses, in
addition to
the superconformal symmetry described by the Lie
superalgebra
$osp(2|2)$, a hidden nonlinear symmetry  $osp(2|2)_n$.
\begin{itemize}
\item
It would be interesting to trace out a manifestation of
this additional symmetry
as well as of the nonlinearly generalized superconformal
symmetry of the fermion-monopole system
in the context
of the corresponding quantum scattering problems.
\end{itemize}
By analogy with the anyon models,
the shift of the angular momentum
of the free planar fermion system may be interpreted
as proceeding from the coupling of the particle
to the magnetic field of the singular magnetic flux.
\begin{itemize}
\item
Therefore, the analysis of the symmetries of the
planar model of a charged particle in the field of magnetic
vortex,
which is closely related
to the fermion-monopole system \cite{LP3},
could give a new perspective on the nonlinear superconformal
symmetry $osp(2|2)_n$.
\end{itemize}
As in the case of nonlinear holomorphic
supersymmetry \cite{KP2}, the
nonlinear superconformal symmetry may be treated
as a symmetry of a higher spin particle system \cite{AnPl}.
Proceeding from the  close similarity between the
fermion-monopole system and superconformal mechanics model,
\begin{itemize}
\item
one could expect the appearance of
some generalization of the nonlinear superconformal
symmetry $osp(2|2)_n$
as a symmetry for a higher spin charged particle in the
field of the Dirac monopole.
\end{itemize}
Due to the observed very close similarity between
the superconformal mechanics and the fermion-monopole
models,
\begin{itemize}
\item
it would also interesting to investigate
the superposition of the both systems
from the point of view of the nonlinear
superconformal symmetry.
\end{itemize}
\vskip 0.3cm
{\bf Acknowledgements.}
I thank
E. Ivanov, S. Krivonos and A. Pashnev for useful
and stimulating discussions.
This work was partially supported by the grant
7010073 from FONDECYT (Chile) and by DICYT-USACH.


\begin{thebibliography}{99}

\bibitem{AIS}
A.~A.~Andrianov, M.~V.~Ioffe and V.~P.~Spiridonov,
Phys.\ Lett.\ A {\bf 174} (1993) 273
[hep-th/9303005];
A.~A.~Andrianov, M.~V.~Ioffe and D.~N.~Nishnianidze,
Phys.\ Lett.\ A {\bf 201} (1995) 103
[hep-th/9404120].

\bibitem{P1}
M.~Plyushchay,
Int.\ J.\ Mod.\ Phys.\ A {\bf 15} (2000) 3679
[hep-th/9903130].

\bibitem{KP1}
S.~M.~Klishevich and M.~S.~Plyushchay,
Nucl.\ Phys.\ B {\bf 606} (2001) 583
[hep-th/0012023].

\bibitem{A1}
H. Aoyama, M. Sato and T. Tanaka,
Phys. Lett. B {\bf 503} (2001) 423
[quant-ph/0012065];
Nucl.\ Phys.\ B {\bf 619} (2001) 105
[quant-ph/0106037].


\bibitem{Walg}
J. de Boer, F. Harmsze, T. Tjin,
\textit{Phys. Rept.} \textbf{272} (1996) 139
[hep-th/9503161].


\bibitem{KP3}
S.~M.~Klishevich and M.~S.~Plyushchay,
Nucl.\ Phys.\ B {\bf 628} (2002) 217
[hep-th/0112158].

\bibitem{AP}
V.~P.~Akulov and A.~I.~Pashnev,
Teor.\ Mat.\ Fiz.\  {\bf 56} (1983) 344.

\bibitem{FR}
S.~Fubini and E.~Rabinovici,
Nucl.\ Phys.\ B {\bf 245} (1984) 17.

\bibitem{IKL}
E.~A.~Ivanov, S.~O.~Krivonos and V.~M.~Leviant,
J.\ Phys.\ A {\bf 22} (1989) 4201.

\bibitem{LP1}
C.~Leiva and M.~S.~Plyushchay,
JHEP 0310 (2003) 069
[hep-th/0304257].

\bibitem{AnPl}
A.~Anabalon and M.~S.~Plyushchay,
Phys.\ Lett.\ B {\bf 572} (2003) 202
[hep-th/0306210].

\bibitem{LP2}
C.~Leiva and M.~S.~Plyushchay,
Phys. Lett. B (in press),
arXiv:hep-th/0311150.

\bibitem{Jackiw}
R.~Jackiw,
Annals Phys.\  {\bf 129} (1980) 183.

\bibitem{mpmono}
M.~S.~Plyushchay,
Nucl.\ Phys.\ B {\bf 589} (2000) 413
[hep-th/0004032].


\bibitem{DHV}
E.~D'Hoker and L.~Vinet,
Phys.\ Lett.\ B {\bf 137} (1984) 72.

\bibitem{MacHol}
F.~De Jonghe, A.~J.~Macfarlane, K.~Peeters and J.~W.~van
Holten,
Phys.\ Lett.\ B {\bf 359} (1995) 114
[hep-th/9507046].

\bibitem{Spec}
D.~Spector,
Phys.\ Lett.\ B {\bf 474} (2000) 331
[hep-th/0001008].

\bibitem{PlFM}
M.~S.~Plyushchay,
Phys.\ Lett.\ B {\bf 485} (2000) 187
[hep-th/0005122].

\bibitem{Horv}
P.~A.~Horvathy, A.~J.~Macfarlane and J.~W.~van Holten,
Phys.\ Lett.\ B {\bf 486} (2000) 346
[hep-th/0006118].


\bibitem{KP2}
S.~M.~Klishevich and M.~S.~Plyushchay,
Nucl.\ Phys.\ B {\bf 616} (2001) 403
[hep-th/0105135];
Nucl.\ Phys.\ B {\bf 640} (2002) 481
[hep-th/0202077].

\bibitem{LP3}
C.~Leiva and M.~S.~Plyushchay,
Annals Phys.\  {\bf 307} (2003) 372
[hep-th/0301244].




\end{thebibliography}
\end{document}